\documentclass[useAMS,usenatbib,psfig]{mn2e}

\usepackage{graphicx}

\title[Ab initio simulations of accretion disks instability]
{Ab initio simulations of accretion disks instability}
\author[V. Teresi, D. Molteni, E. Toscano]{V. Teresi$^1$ \thanks{E-mail:
vteresi@unipa.it (VT)}, D. Molteni$^1$ and E. Toscano
$^{1}$ \\
$^{1}$Dipartimento di Fisica e Tecnologie Relative, Universit$\grave{a}$ di Palermo,
Viale delle Scienze, Palermo, 90128, Italy\\
}
\begin{document}



\maketitle

\label{firstpage}

\begin{abstract}
We show that accretion disks, both in the subcritical and
supercritical accretion rate regime, may exhibit significant
amplitude luminosity oscillations. The luminosity time behavior
has been obtained by performing a set of time-dependent 2D SPH
simulations of accretion disks with different values of $\alpha$
and accretion rate. In this study, to avoid any influence of the
initial disk configuration, we produced the disks injecting matter
from an outer edge far from the central object. The period of
oscillations is $2-50$ s respectively for the two cases, and the
variation amplitude of the disc luminosity is $10^{38}$ -
$10^{39}$ erg/s. An explanation of this luminosity behavior is
proposed in terms of limit cycle instability: the disk oscillates
between a radiation pressure dominated configuration (with a high
luminosity value) and a gas pressure dominated one (with a low
luminosity value). The origin of this instability is the
difference between the heat produced by viscosity and the energy
emitted as radiation from the disk surface (the well-known thermal
instability mechanism). We support this hypothesis showing that
the limit cycle behavior produces a sequence of collapsing and
refilling states of the innermost disk region.
\end{abstract}


\section{Introduction}

This work continues our studies on the occurrence of the Shakura
and Sunyaev instability \citep{b12}  in the
$\alpha$-disks when the radiation pressure dominates, i.e. in the so-called A zone.\\
The problem of the existence and outcome of the Shakura-Sunyaev
instability is important in accretion disc physics because it
affects the models and their time behavior.\\
In general, the outcome of the Shakura-Sunyaev instability is
guessed to be the formation of a hot cloud around
the internal disc region, in which comptonization could happen \citep{b14}.\\
A critical point of this scenario is the typical timescale
required by the disk to leave the collapsed 'dead' state. Time
dependent analytical models of the disk evolution in this post
collapsed phase are very difficult. Numerical simulations are
therefore important and essential tools to obtain some indication
of the outcome of this evolution.\\
Recently some authors have investigated this problem through the
numerical approach. Szuszkiewicz and
Miller \citep{SM1997} found that a slim accretion disc model with low
viscosity ($\alpha = 0.001$) and a luminosity higher than $0.08
L_E$ shows a thermal instability which gives rise to a shock-like
structure near to the sonic point, leading to the disc disruption.
They found no limit-cycle behavior, probably, according to their
own conclusions, because of the not strong enough advection. The
same Szuszkiewicz and Miller \citep{SM1998} also performed numerical
simulations of accretion disc models with high viscosity ($\alpha
= 0.1$) and obtained a limit-cycle behavior. They simulated the
disc evolution for several cycles and, for $\dot{M}=0.06\dot{M}_E$
and a central object of $10$ solar masses, found a period of the
cycle of about $780$ s. In both papers they reported the results
concerning a vertically integrated disc model, with no
consideration of acceleration in the vertical direction. The same
authors \citep{SM2001} performed finally a
numerical study of an accretion disc model at high viscosity
($\alpha = 0.1$) with a vertically integrated treatment of
acceleration in the vertical direction and a diffusive form for
the viscosity instead of the $\alpha P$ prescription used in their
previous works. Also with this more refined model they found a
limit-cycle behavior.\\
Nayakshin et al. \citep{Nayakshin} used a limit-cycle model to
explain the luminosity variability of the micro-quasar GRS
1915+105. Their model is different from the one used by
Szuszkiewicz and Miller. The essential difference regards the
viscosity prescription. Szuszkiewicz and Miller used the standard
Shakura-Sunyaev one or the more refined (but fundamentally
equivalent) diffusive formulation. In their 1D simulations the
discs oscillate between two stable states, one at high luminosity
and the other with a much lower emission. These two states are the
standard gas-pressure dominated one and the radiation pressure
dominated one (note that this last state is stable in the slim
accretion disc model). If the accretion rate difference between
the high and low states is very large, as in GRS 1915+105, the
high state should last a very short time. But in reality the GRS
1915+105 has a high state lasting for a long time, even more than
the low state. Therefore the limit-cycle model in slim accretion
discs cannot explain the time behavior of this source. So
Nayakshin et al. used a particular viscosity law which produces a
high stable state of larger duration. With this model, they
explained the gross observational features of GRS 1915+105.\\
Janiuk et al. \citep{Janiuk} also tried to describe the GRS
1915+105 behavior in terms of a limit-cycle model. They adopted
the standard $\alpha$-viscosity prescription, but included in the
model the effect of a corona surrounding the disc and a vertical
outflow. With half of the energy dissipated in the disc, they
obtained outbursts whose amplitude and duration
are consistent with the GRS 1915+105 data.\\
Teresi et al. \citep{ter2003} have clearly shown that, at intermediate
accretion rates, accretion discs with A zone suffer a collapse but
after a rather long time they show a flaring activity with an
intervening refilling phase of the A zone.\\
We point out that our simulations differ from the Szuszkiewicz and Miller ones since we
produce real 2D disks with true vertical motion. No 'ad hoc'
prescription is required to include physical vertical effects. The
new degree of freedom given by the Z motion has many consequences,
which we discuss in section 4.


Furthermore we point out that all previous simulations start from
a {\it full disk existing at time $t=0$}. A typical drawback of
simulations involving large disk sectors is the uncertainty in the
initial model structure. Indeed the analytical models lack a
reliable vertical structure \citep{b3,Hubeny}. The disk luminosity
and other disk parameters can oscillate for long time before
reaching a steady configuration and it can be difficult to discern
between real instability oscillations and transient spurious
oscillations, that could last a long time.

The simulations we show here differ from our previous ones and
from many other authors since the disc is generated "ab initio".
We inject matter at a large distance from the central object. The
injected gas has low temperature and keplerian angular momentum.
Its evolution is due to the action of the viscous stress, whose
$\alpha$ value is given. In this way the disk evolves smoothly
through a series of equilibrium states, avoiding the problem of the
transient spurious oscillations and the influence of the initial
configuration on the final result.
It is clear that also for these
simulations  a long integration time (of the order of the  viscous
drift time) is required. The numerical Smoothed Particles
Hydrodynamics technique allows to integrate up to so long times.
Let us note
that -in general- a lagrangean code, as SPH is, is better suited to capture
convective motions than eulerian codes. With the same spatial
accuracy (cell size equal to particle size) the SPH particle
motion is tracked with great accuracy, i.e. the particle size may
be large but its trajectory can be still determined 'exactly'.

Our results suggest that the Shakura-Sunyaev model can be used to
explain the luminosity variability shown by many sources.
The aim of this work, however, is not to find an explanation of some sources behavior, but
simply to see what happens to the standard disc structure during
its time evolution.

The paper is structured as follows: in section 2 we remind the
physical model, in section 3 we describe the adopted numerical
method, in section 4 we report on the simulated cases and the
obtained results, presenting and commenting some figures, in
section 5 the physical aspects of the simulations results are
discussed and in section 6 we expose conclusions and astrophysical
implications of our work.

\section[]{The physical model}
The time dependent equations describing the physics of accretion
disks are well known. We used the lagrangean form of them in a
cylindrical reference system and in the approximation of local
thermal equilibrium (LTE) between gas and radiation
\citep{mihalas}.
They include: \\
mass conservation

\begin{equation}
\frac{D\rho}{Dt} = -\rho \hspace{0.20cm}
div\stackrel{\rightarrow}{v}
\end{equation}

radial momentum conservation

\begin{equation}
\rho \frac{Dv_r}{Dt} = -\rho \frac{\lambda^2}{r^3} + \rho g_r + (div
\stackrel{\rightarrow }{\stackrel{\rightarrow }{\sigma }})_r + f_r
\end{equation}

vertical momentum equation

\begin{equation}
\frac{Dv_z}{Dt} = -\frac{1}{\rho} \frac{dP}{dz} - g_z + \frac {f_z}{\rho}
\end{equation}



energy equation

$$
\frac D{Dt}\left( \frac{E_{rad}}\rho +\epsilon +\frac 12v^2\right)
=\stackrel{\rightarrow }{v}\cdot \stackrel{\rightarrow
}{g}-\frac{\left( P_{rad}+P_{gas}\right) }\rho \nabla
\stackrel{\rightarrow }{v}+\stackrel{%
\rightarrow }{v}\cdot \frac {\stackrel{\rightarrow }{f}}{\rho}
$$
\begin{equation}
+\frac 1\rho \nabla \left[
\stackrel{\rightarrow }{v}\stackrel{\rightarrow }{\stackrel{\rightarrow }{%
\sigma }}\right]
\end{equation}
where $\sigma _{ij}$ , $\stackrel{\rightarrow
}{\stackrel{\rightarrow }{\sigma }} $, is the viscosity stress
tensor and $\stackrel{\rightarrow }{g}$ is the body force per unit
mass (acceleration).

angular momentum equation

\begin{equation}
\frac{D\Omega}{Dt} = -2\Omega\frac{v_r}{r} +
\frac{1}{\rho}\frac{\partial}{\partial z}(\nu \rho \frac{\partial
\Omega}{\partial z}) +
\frac{1}{\rho}\frac{1}{r^3}\frac{\partial}{\partial r}(r^3 \nu
\rho \frac{\partial \Omega}{\partial r})
\label{diff}
\end{equation}

Here $\Omega$ is the local angular velocity, $\frac{D}{Dt}$ is the
comoving derivative, $E_{rad}$ is the radiation energy per unit
volume, $\stackrel{\rightarrow}{f}$ is the radiation force per
unit volume, given by:

\begin{equation}
\stackrel{\rightarrow}{f} = \rho \frac{k + \sigma_T}{c}
\stackrel{\rightarrow}{F}
\end{equation}

$\stackrel{\rightarrow}{F}$ is the radiation flux, given by:

\begin{equation}
\stackrel{\rightarrow}{F} = - \frac{c}{3\rho(k + \sigma_T)}
\stackrel{\rightarrow}{\nabla}E_{rad}
\end{equation}

$k$ and $\sigma_T$ are the free-free absorption and Thomson
scattering coefficients, given by:

\begin{equation}
k = c_k \rho T^{-\frac{7}{2}}    [cm^2 g^{-1}]
\end{equation}

(Kramers' formula, valid since our temperatures are well beyond
$10^{4}$ K)

with $c_k = 2.26 \cdot 10^{24}$, and

\begin{equation}
\sigma_T = 0.4    [cm^2 g^{-1}]
\end{equation}

$\lambda$ is the angular momentum per unit mass, $E = \epsilon +
\frac{E_{rad}}{\rho}$ is the total internal energy per unit mass,
including gas and radiation terms.\\
\\
The components of $\stackrel{\rightarrow }{\stackrel{\rightarrow
}{\sigma }} $ that we have considered in our calculations (since
they are the ones that play an important role in accretion discs)
are the $r$-$\phi$ one, given by:

\begin{equation}
{\sigma}_{r\phi} = \nu \rho r \frac{\partial\Omega}{\partial r}
\end{equation}

and the $\phi$-$z$ one, given by:

\begin{equation}
{\sigma}_{\phi z} = \nu \rho \frac{\partial(\Omega r)}{\partial z}
\end{equation}

$\nu = \alpha v_s H$ is the kinematic viscosity, $\alpha$ is the
viscosity parameter of the Shakura-Sunyaev model, $v_s$ is the
local sound speed, $H$ is the disc vertical thickness and the other
terms have the usual gas dynamic meaning.\\
The gravitational field produced by the black hole is given by the
well-known pseudo-newtonian formula by \citet{b10}:

\begin{equation}
\vec{E}_{grav} = -\frac{GM}{(R-R_{g})^{2}}\frac{\vec{R}}{R}
\end{equation}

where $\vec{R}$ is the position vector of the point in which the
field is evaluated, with a modulus given by $R = \sqrt{r^2 +
z^2}$, $R_g$ is the Schwartzschild gravitational radius of the
black hole, given by:

\begin{equation}
R_g = \frac{2GM}{c^2}
\end{equation}

and $M$ is the black hole mass.\\
\\
We adopt the local thermal equilibrium approximation for the
radiation transfer treatment. However this assumption does not
affect our conclusions.

\section{The numerical method}
We set up a new version of the Smoothed Particles Hydrodynamics
(SPH) code in cylindrical coordinates, for axis symmetric
problems. We remind that SPH is a lagrangean interpolating method.
Recently it has been shown it is equivalent to finite elements
with sparse grid nodes moving along the fluid flow lines
\citep{Dilts}. For a detailed account of the SPH algorithm see
\citet{b9}. For cylindrical coordinates implementation see
\citet{moltbook} and \citet{moltchakra}. Our code includes
viscosity and radiation treatment.

The basic point for our cylindrical geometry approach is simply to
assume a usual kernel function but depending directly on the
radial (r) and vertical (z) variables, and therefore retaining the
usual normalization factor and width. Now pseudoparticles are
small tori of mass $dm_k=2\pi\varrho _kr_kdr_kdz_k$ . In this way
we may use the same Cartesian grid in the $(r,z)$ domain and the
same procedure to search the near neighbors of each particle.
Therefore
applying the usual procedure 
for the evaluation of any smooth function in the point $(r_i,z_i)$ we have:
$$
{f({\bf r}_i)}={\int_Vf({\bf r^{\prime }})W_h({\bf r}_i-{\bf
r^{\prime }}) \frac{2\pi r^{\prime } \rho({\bf r^{\prime }})}{2\pi
r^{\prime } \rho({\bf r^{\prime }})} d{\bf r^{\prime }}}\simeq
$$
\begin{equation}
\simeq \sum_{j=1}^Nf({\bf r}_j){\ \frac{m_j}{2\pi
r_j\rho_j}}W_{ij},
\end{equation}
where ${\bf r}_k=(r_k,z_k)$. So for the density we have the simple
expression that identically satisfies the continuity equation in
the cylindrical form:
\begin{equation}
\rho({\bf r}_i)\simeq \sum_{j=1}^N\frac{m_j}{r_j}W_{ij}
\end{equation}
Rewriting the fundamental equations in the formulation more
suitable for the SPH evaluation \citep{b9}, and applying the
previous criteria we have the following expressions; for the
radial (r) momentum we obtain:
\begin{equation}
\left( \frac{Dv_r}{Dt}\right) _i=-\frac{v_\phi ^2}r-\sum_{j=1}^N\frac{m_j}{%
r_j}\left( \frac{p_i}{\rho_i^2}+\frac{p_j}{\rho_j^2}+\Pi _{ij}\right) \frac{%
\partial W_{ij}}{\partial r_i}~~~.
\end{equation}
where $\Pi$ is the artificial viscosity pressure.\\
\\
The vertical (z) momentum satisfies:
\begin{equation}
\left( \frac{Dv_z}{Dt}\right) _i=-\sum_{j=1}^N\frac{m_j}{r_j}\left( \frac{%
p_i }{\varrho _i^2}+\frac{p_j}{\varrho _j^2}+\Pi _{ij}\right) \frac{\partial
W_{ij}}{\partial z_i}~~~.
\end{equation}



For the energy equation we based our implementation on the following procedure.
Let us call $U=\frac{E_{rad}}\rho
+\epsilon +\frac 12v^2$ , $P^{tot}=P_{rad}+P_{gas}$ ,and remind that  $\stackrel{%
\rightarrow }{F}=-\frac 1\rho \nabla \left( P_{rad}+P_{gas}\right)
$ is the total force per unit mass due to gas and radiation, then
the first three terms of the energy formula can be trivially put
into the SPH formalism according to standard prescriptions
\citep{mon1992}. So we obtain:
$$
\frac{dU_i}{dt}=\left( \stackrel{\rightarrow }{v}\cdot
\stackrel{\rightarrow }{g}+\stackrel{\rightarrow }{v}\cdot
\stackrel{\rightarrow }{F}\right) _i+\sum_{k=1}^Nm_k\left(
\frac{P_i^{tot}}{\rho _i^2}+\frac{P_k^{tot}}{\rho _k^2}\right)
\stackrel{\rightarrow }{v_{ik}}\cdot \nabla _iW_{ik}
$$
\begin{equation}
+\left[
\frac 1\rho \nabla \left( \stackrel{\rightarrow }{v}\cdot \stackrel{%
\rightarrow }{\stackrel{\rightarrow }{\sigma }}\right) \right] _i
\end{equation}
here  $\stackrel{\rightarrow }{v_{ik}}=\stackrel{\rightarrow }{v_i}-%
\stackrel{\rightarrow }{v_k}$. The fourth term $\left[ \frac 1\rho \nabla
\left( \stackrel{\rightarrow }{v}\stackrel{\rightarrow }{\stackrel{%
\rightarrow }{\sigma }}\right) \right] _i$ can be written, using
the same method of SPH evaluation for the $\nabla
\stackrel{\rightarrow }{v}$ term of the continuity equation, as:
\begin{equation}
\left[ \frac 1\rho \nabla \left( \stackrel{\rightarrow }{v}\cdot \stackrel{%
\rightarrow }{\stackrel{\rightarrow }{\sigma }}\right) \right] _i=\sum_{k=1}^N
\frac{m_k}{\left( \frac{\rho _i+\rho _k}2\right) }\stackrel{\rightarrow }{%
S_{ik}}\cdot \nabla _iW_{ik}
\end{equation}
where we have symmetrized the density term, and where
$\stackrel{\rightarrow }{S}=\stackrel{\rightarrow }{v}\cdot \stackrel{%
\rightarrow }{\stackrel{\rightarrow }{\sigma }}$,
$\stackrel{\rightarrow }{S_{ik}}=\stackrel{\rightarrow
}{S}_i-\stackrel{\rightarrow }{S}_k$; with this procedure the SPH
energy equation conserves exactly the total energy.

The total thermal internal energy $\frac{E_{rad}}\rho +\epsilon $
is recovered by subtraction of the kinetic energy and then the
ratio between $\frac{E_{rad}}\rho $ and $\epsilon $ is given
requiring the LTE condition.

In cylindrical coordinates the particles masses $m_i$ must be replaced by $
\frac{m_i}{r_i}$ and a further $r_k$ term appears in the terms coming out of
the divergence expressions, so we have for example:%
$$
\sum_{k=1}^Nm_k\left( \frac{P_i^{tot}}{\rho _i^2}+\frac{P_k^{tot}}{\rho _k^2}%
\right) \stackrel{\rightarrow }{v_{ik}}\cdot \nabla _iW_{ik}\Rightarrow
$$
$$
\Rightarrow \sum_{k=1}^N\frac{m_k}{r_k}\left(
\frac{P_i^{tot}}{\rho _i^2}+\frac{P_k^{tot} }{\rho _k^2}\right)
\stackrel{\rightarrow }{V}_{ik}^{cyl}\cdot \nabla _iW_{ik}
$$
where $\stackrel{\rightarrow }{V}_{ik}^{cyl}=\left(
r_iv_{r_i}-r_kv_{r_k}\right) \stackrel{\symbol{94} }{r}+\left(
v_{z_i}-v_{z_k}\right) \stackrel{\symbol{94} }{z}$ , and $\stackrel{\symbol{%
94} }{r}$ and $\stackrel{\symbol{94} }{z}$ are the radial and Z versors.

To derive all previous expressions we neglect the contributions to
the integrals from the boundary of the integration domain. The
artificial viscosity pressure $\Pi _{ij}$ is formulated as
\begin{equation}
\Pi _{ij}=\frac{\alpha \tilde{\mu}_{ij}{\bar{c}}_{ij}+\beta \tilde{\mu}%
_{ij}^2}{{\bar{\rho}}_{ij}}~~~,
\end{equation}
with the averaged quantities
$$
{\bar{c}}_{ij}=\frac{c_i+c_j}2~~,~~~{{\bar{\rho}}_{ij}}=\frac{\rho _i+\rho _j%
}2~~,
$$
$$
~~~\tilde{\mu}_{ij}=\frac{r_iv_{ri}-r_jv_{rj}}{r_i(l_{ij}^2+\eta
^2)}+ \frac{(v_{zi}-v_{zj})(z_i-z_j)}{(l_{ij}^2+\eta ^2)}~~,
$$
$$
l_{ij}^2=(r_i-r_j)^2+(z_i-z_j)^2~~,~~~\eta =0.1h~~,
$$
$\alpha $ and $\beta $ are the artificial viscosity coefficients used to
damp out oscillations in shock transitions, $c$ here denotes the sound speed.

Since our aim is to simulate accretion discs it was essential a
correct treatment of the tangential velocity and its diffusion due
to the viscosity. We integrate explicitly the viscosity diffusion
term; the cylindrical SPH-version of the diffusion term of
equation (\ref{diff}) is given following the criteria by
\citet{brookshaw}:

\begin{equation}
\left( \frac{\partial \Omega }{\partial t}\right) _i=\sum_{j=1}^N\frac{m_j}{%
r_j}\left( \frac{\Omega _i-\Omega _j}{\varrho _i\varrho _j}\right) D_{ij}
\frac{{\bf R}_{ij}}{R_{ij}^2}\cdot \nabla _iW_{ij}~~~,
\end{equation}
where
\begin{equation}
D_{ij}=\frac{\mu _ir_i^3+\mu _jr_j^3}{r_i^3}\ \ ,\ \ \ {\bf R}%
_{ij}=(r_i-r_j,z_i-z_j)
\end{equation}

The formulae for cylindrical geometry are similar to the cartesian
ones and the most relevant changes are : (i) the mass of a
particle appears divided by its distance from the z-axis (ii)
mutual velocity difference $v_j-v_i$
between two particles must be replaced by the more sophisticated term $({r_i%
{\bf v}_i-r_j{\bf v}_j})/{r_i}$.

The force $F_{r_{ji}}$ differs from $F_{r_{ij}}$ while $%
F_{z_{ij}}=F_{z_{ji}} $ for particles at the same radial coordinate. This
difference in the force is due to the geometry.
Angular momentum is exactly conserved. The statement $dm_k=2\pi
\varrho _kr_kdr_kdz_k$ is needed only for the derivation of the formulae and
the particles in the simulations may have the same mass or not; obviously
the density is no more directly proportional to the number of particles per
unit area as for the case of particles having the same mass.

To integrate the energy equation we adopted
the splitting procedure. In the LTE condition the radiation energy
density changes according to the well known diffusion equation
given by:

\begin{equation}
\frac{\partial E_{rad}}{\partial t}=
-div\stackrel{\rightarrow}{F}= \stackrel{\rightarrow}{\nabla}
\cdot \left( \frac c{3\rho \kappa_{tot}
}\stackrel{\rightarrow}{\nabla} E_{rad}\right)
\end{equation}

where $k_{tot} = k + \sigma_T$.

In cylindrical coordinates $r,z$:

\begin{equation}
\frac{\partial E_{rad}}{\partial t}=\frac cr\frac \partial {\partial r}\left(
\frac r{3\rho \kappa_{tot} }\frac{\partial E_{rad}}{\partial r}\right) +\frac cr\frac
\partial {\partial z}\left( \frac r{3\rho \kappa_{tot} }\frac{\partial E_{rad}}{%
\partial z}\right)
\end{equation}

where $c$ is the light speed.\\
\\

The SPH-version of the radiation transfer term is given following
the criteria by \citet{brookshaw}. The cylindrical coordinate
version is given by:

\begin{equation}
\left( \frac{\partial E}{\partial t}\right) _i=\frac 1{r_i}\sum_{j=1}^N\frac{%
m_j}{r_j}\left( \frac{E_i-E_j}{\rho _j}\right) D_{ij}\frac{{\bf R}_{ij}}{%
R_{ij}^2}\cdot \stackrel{\rightarrow}{\nabla} _iW_{ij}
\end{equation}

where for clarity we did not put the subscript $rad$ in $E_{rad}$
and where:

\begin{equation}
D_{ij}=\left( \frac{cr_i}{3\rho _i\kappa _{tot_i}}+\frac{cr_j}{3\rho _j\kappa _{tot_j}}%
\right) \ \ ,\ \ \ {\bf R}_{ij}=(r_i-r_j,z_i-z_j)
\end{equation}

This formula can be obtained by the same procedure explained by
Brookshaw, but taking into account that -in cylindrical
coordinates- the particles masses are defined as $m_k=2\pi \ \rho
_k\ r_k\ \Delta r_k\ \Delta z_k$, that explains the further
division by $r_j$ in the term $\frac{m_j}{r_j}$ .

We used a variable $h$ procedure \citep{hvar}.
In our procedure, in order to
have a not too small particle size (and therefore not too great
CPU integration times), we put a floor for the $h$ values: $h$ is
chosen as the maximum between the value given by the variable $h$
procedure itself and $1/10$ of the disc vertical half thickness.
So we have nearly 10 particles along the disc half thickness even
in the collapsed region.

The boundary conditions of the simulations are not fixed, though
we produce an inflow at a certain radius, generating new particles
with fixed density and temperature every time a circular zone
around the injector position becomes empty.
As the SPH particles move around, the simulation region follows the form
assumed by the disc and the values of the physical variables at
the boundary of the disc are the values that
characterize the boundary particles at a certain time.\\
The spatial extension of the initial configuration is decided
by establishing a radial range of physical interest and a vertical
extension given by the disc thickness of the Shakura-Sunyaev model.
The values are given in the next section.\\
For radiation, the boundary conditions we used are based on the
assumption of the black-body emission and particularly on the
Brookshaw approximation \citep{brookshaw}. At every time step
boundary particles are identified by geometrical criteria (the
particle having the maximum absolute value of $z$ in a vertical
strip of radial width given by $h$ is a boundary particle). The
boundary particle loses its thermal energy according to the
formula given by Brookshaw (that is an approximation of the
diffusion equation at the single particle level), that states the
particle cooling rate proportional to $(QT)/h^2$, where $Q =
(4acT^3)/(3\rho \kappa_{tot})$.

In all our simulations the boundary particles never reach an
optical thickness lower than 10.

\section{The simulations performed}

We performed several simulations, the ones commented here had the
following parameter values:

a)     $\alpha = 0.1$,    $\dot{M} = 0.15$,  domain $R_{1}-R_{2}
= 3-100$, $h = 0.25$;

b)     $\alpha = 0.1$,    $\dot{M} = 2$,  domain $R_{1}-R_{2}
= 3-200$, $h = 0.5$;

where $\dot{M}$ is in units of $\dot{M}_E$ and $\dot{M}_E$ is the
critical accretion rate. For all cases the central black hole mass
is $M = 10 \ M_{\odot}$. The initial spatial resolution we adopt is $h$.\\
The reference units we use are $R_g$ for length values, $R_g/c$
for time values and $L_{theor} = 0.06 \dot{M} c^2$, the
theoretical luminosity for an accretion disc around a not rotating
black hole, for the luminosity values. In case 'a' $L_{theor} =
1.54 \cdot 10^{38} \hspace{0.20cm} erg \hspace{0.20cm} sec^{-1}$;
in case 'b' $L_{theor} = 2.05 \cdot 10^{39} \hspace{0.20cm} erg
\hspace{0.20cm} sec^{-1}$. We have chosen these units because the
simulations results obtained with certain values of the parameters
$M$, $\dot{M}$ and $\alpha$, if given in terms of adimensional
units, can be easily generalized to other systems.\\
The radial domain of each simulation has been chosen with the aim
of including in the simulation a sufficiently wide portion of the
radiation dominated zone, the so-called A zone (in the case 'a' the
entire A zone is included in the simulation).
The $h$ values above reported are the initial ones. They have been
chosen in order to have a good spatial resolution at the injection
radius. The variable $h$ procedure guarantees then an equally good
resolution in the inner disc regions.

For case 'a' we stress that our results have been obtained
simulating a full disc including A, radiation dominated, and B,
gas pressure dominated, zones. The presence of the B zone, that is
theoretically stable and that we see stable in our simulations,
guarantees in general the numerical accuracy of our study and
allows to clearly identify the A zone as responsible of the
oscillations.

In this section we want to show the changes that occur in the main
properties and physical quantities of the disc due to the
instability and the consequent limit-cycle behavior. When the
instability arises the disc undergoes a collapse phase, with a
strong lowering of its vertical thickness. Fig. 1 makes evident
the effect of this phenomenon, showing, for the case 'a', the disc
configuration reached at the end of the collapse phase,
characterized by a very small Z height in the innermost region ($r
< 13R_g$). Note that the Z scale is graphically amplified in the
figure.

\begin{figure}
\begin{center}
\includegraphics[scale = 0.33,angle = 270.]{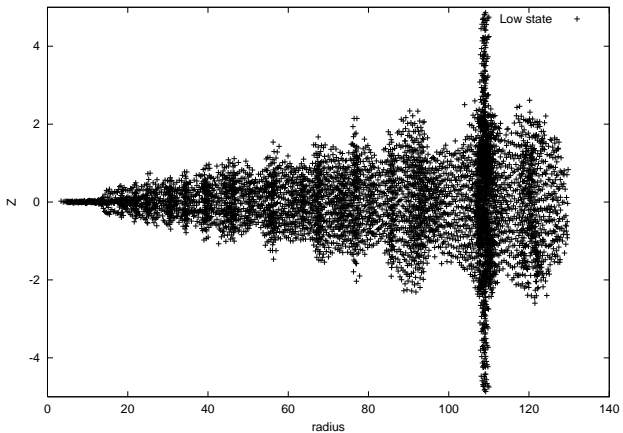}
\caption{The $r$-$z$ profile of the disc of case 'a' in its low
state is shown at the time $t = 0.52155 \ 10^7 \ R_g/c$ . Every
SPH particle is represented by a small cross. On the $x$ axis the
$r$ values in units of $R_g$ are represented. On the $y$ axis the
$z$ values in units of $R_g$ are represented.}
\end{center}
\end{figure}

In this state the mass accretion rate is no more uniform
throughout the whole disc. In fact, in the innermost, collapsed
disc zone $\dot{M}$ has a value lower than the one before the
collapse, whereas at the outer boundary of this zone it assumes
the value of the outer, not collapsed region, i.e. the unperturbed
value. So mass is forced to enter the collapsed zone at a rate
larger than the one at which mass falls into the black hole. Due
to this accretion rates difference, the collapsed disc zone is
refilled and consequently reaches a configuration of much larger
vertical thickness (comparable to the one of the
unperturbed state). In fig. 2 we show the $r$-$z$ profile of this new structure,
together with the boundaries of the corresponding Shakura-Sunyaev disc,
determined by calculating the disc vertical thickness (at all the $r$ values inside
the simulation radial range) from the Shakura-Sunyaev one-dimensional model
with the same accretion rate and $\alpha$. This comparison of the 2D simulated
model and the 1D theoretical one is shown in order to make evident the agreement
we obtained between the results of the simulations and the canonical disc model.
We will discuss more deeply this point in the section 5.

\begin{figure}
\begin{center}
\includegraphics[scale = 0.33,angle = 270.]{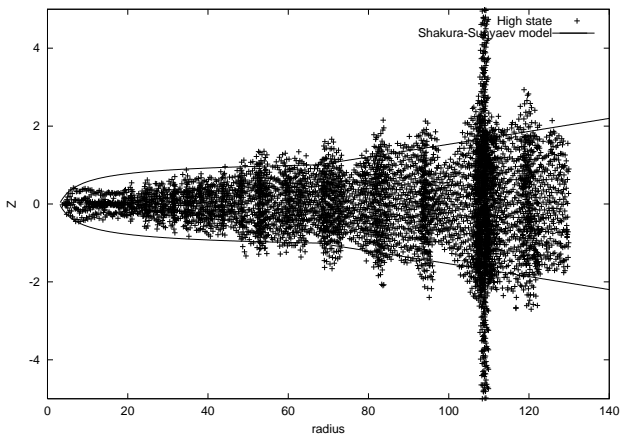}
\caption{The $r$-$z$ profile of the same disc case 'a' in its high
state at the time $t = 0.52233 \ 10^7 \ R_g/c$ is shown. The solid
line represents the equivalent Shakura-Sunyaev model.}
\end{center}
\end{figure}

Moreover, the difference between the two states (collapsed and
refilled) is not only in the value of the disc vertical thickness.
In the unstable disc region the temperature of the collapsed state
is lower than the one of the refilled configuration. As a
consequence of that, the ratio between radiation and gas pressure
is changed from a state to the other one: though in the unstable
region the disc always remains radiation pressure dominated,
during the collapsed state the ratio $P_{rad}/P_{gas}$ is much
lower (close to 1) than in the refilled disc. In fig. 3 we show
the comparison between the radial profiles of the ratio
$P_{rad}/P_{gas}$ in the two states, which makes evident the great
lowering of this ratio in the transition to the collapsed state
within the unstable disc region, and in fig. 4 a similar
comparison for the temperature profiles is shown, making clear
that the unstable region is cooler in the collapsed state than in
the refilled one. In fig. 4 is also shown the temperature radial
profile (in the disc midplane) of the corresponding
Shakura-Sunyaev model. Here also we show the good agreement of our
simulations with the canonical disc model.

\begin{figure}
\begin{center}
\includegraphics[scale = 0.33,angle = 270.]{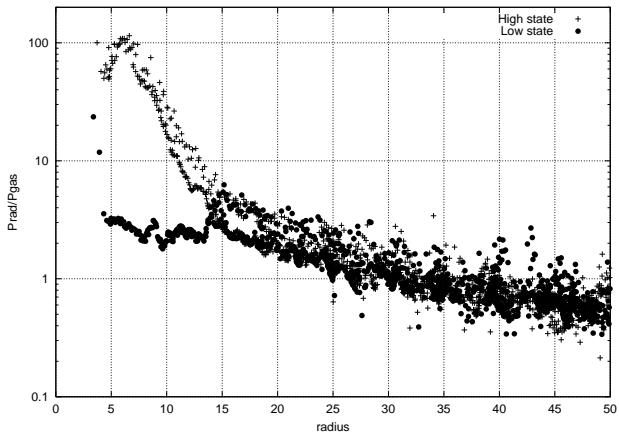}
\caption{The ratio $P_{rad}/P_{gas}$ at the times $t = 0.52155 \ 10^7$ and $t
= 0.52233 \ 10^7$ is shown. On the $x$ axis the $r$ values in units of
$R_g$ are represented. The configuration at the earlier time
exhibits, in the collapsed zone, a much smaller ratio $P_{rad}/P_{gas}$
than the configuration at the later time.}
\end{center}
\end{figure}

\begin{figure}
\begin{center}
\includegraphics[scale = 0.33,angle = 270.]{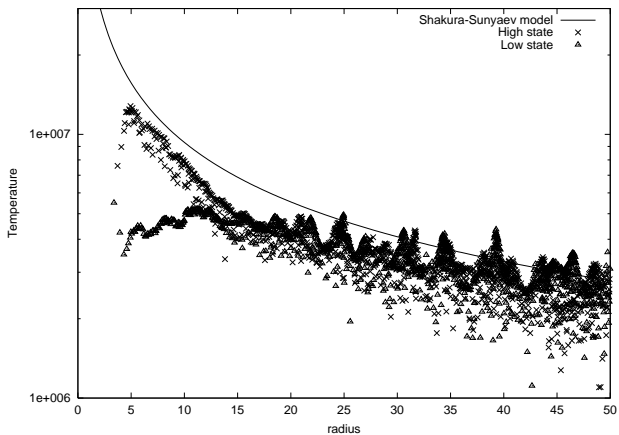}
\caption{The temperature radial profile at the times $t = 0.52155
\ 10^7$ and $t = 0.52233 \ 10^7$ is shown. On the $x$ axis the $r$
values in units of $R_g$ are represented. On the $y$ axis the
temperature values in Kelvin are represented. The configuration at
the earlier time exhibits, in the collapsed zone, a much smaller
temperature than the configuration at the later time. The solid
line represents the temperature radial profile for the equivalent
Shakura-Sunyaev model.}
\end{center}
\end{figure}

In these figures, and in all the figures of this paper in
which physical quantities are plotted versus $r$, we show the values regarding all the
SPH particles: to each particle (of radial coordinate, say, $r$) corresponds in the
figure the point ($r$, $Q$), where $Q$ is the value of the physical quantity that
we are plotting calculated in the particle position.\\
The temperature difference also produces different luminosities
associated to the two configurations. So the disc luminosity
oscillates between the two states and we can observe the
limit-cycle behavior typical of the thermal instability. In fig. 5
we show the time variation of the disc luminosity, from which the
oscillatory behavior is clear. In this figure only a time window
of the luminosity variation regarding the whole history of the
disc is represented. The time units are, as said, $R_g/c$. To
obtain the time values in seconds it is sufficient to multiply the
values in the figure by $10^{-4}$ (the value of $R_g/c$ in seconds
for a black hole of 10 solar masses).

\begin{figure}
\begin{center}
\includegraphics[scale=0.462,angle=0.]{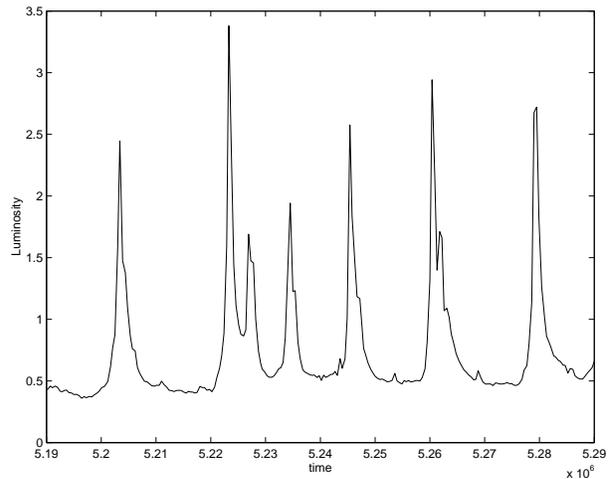}
\caption{The time behavior of the disc luminosity is shown. On the
$x$ axis the time values in units of $R_g/c$ are represented. On
the $y$ axis the luminosity values in units of $L_{theor} = 0.06
\dot{M} c^2 = 1.54 \cdot 10^{38} \hspace{0.20cm} erg
\hspace{0.20cm} sec^{-1}$ are represented.}
\end{center}
\end{figure}

What can be noticed, in particular, from this figure is the shape of the
time variation curve. A single oscillation starts with the disc luminosity $L$ that increases
very steeply; then, after having reached a maximum, $L$ decreases more slowly (with
an exponential-like behavior) until a value close to the initial one is reached.\\
We also evaluated the gas velocity field, finding a significant difference between
the radial speeds ( $V_r$)  in the two states: in the unstable region the refilled disc has a higher
radial speed (with a large spread) than the collapsed one. This is what we can expect considering that the refilled disc is more luminous (since it is hotter) and therefore the accretion rate of
its inner region is larger with respect to the collapsed disc. A larger accretion rate
can be the effect of a larger radial speed. In fig. 6 the radial profiles of  $V_r$
 in the two states are shown.

\begin{figure}
\begin{center}
\includegraphics[scale=0.33,angle=270.]{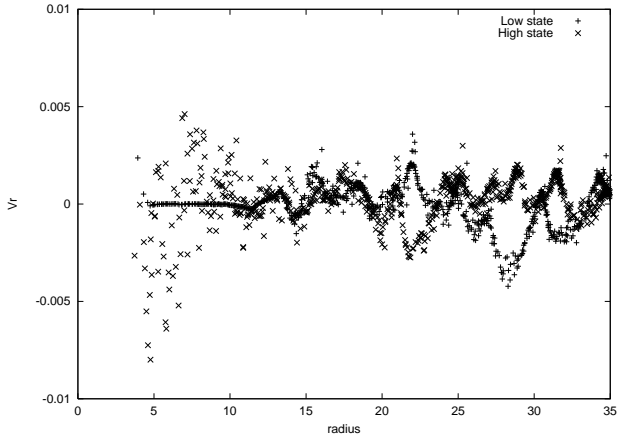}
\caption{The gas radial speed is shown versus $r$ in the two
states (the high and the low one) of the disc. On the
$x$ axis the $r$ values in units of $R_g$ are represented. On
the $y$ axis the radial speed values in units of $c$ are represented.}
\end{center}
\end{figure}

It is evident from this figure what we have said above
and also that in the refilled state the radial speed is often positive, besides very high.
If a large radial speed is present in both inflow and outflow directions, as clearly shown in the figure, we can argue
that there is a gas circulation and not only a large net accretion rate.\\
We highlight that this result excludes the possibility that, in
2D, a Shakura-Sunyaev disc can show a regular radial flow. This
simplified one-dimensional picture is destroyed by the presence of
convective and circulatory motions in the $r-z$ plane. These
motions can be considered as a new turbulence, different from the
one that gives rise to the $\alpha$-viscosity: the equations of
the disc dynamics contain a viscosity term (the $\alpha$-term)
that is physically considered the result of a supposed turbulent
motion, but a disc simulated according to these equations develops
another turbulent motion (that can be supposed to give rise to an
additional viscosity). Our results about circulation and
convection and our hypothesis that these 2D motions can be at the
origin of an additional viscosity agree with the analytical study
by Kippenhahn and Thomas on convective and circulatory flows in
thin radiative accretion discs \citep{Kippen}. The large radial
speed has also to be considered the reason for which the thermal
instability causes the disc collapse and not its expansion. The
local perturbative approach in itself allows to conclude that a
small temperature deviation from the equilibrium state, an
increase as well as a decrease, grows exponentially in time.
Therefore the result should be, with the same probability, an
expansion or a collapse. What we observe, instead, is that
collapse is strongly preferred: in each cycle, initially the inner
zone reduces largely its vertical thickness, then it swells
reaching a thickness value not much larger than the equilibrium
one. Our hypothesis is that what lacks in the local perturbative
approach is the radial drift of matter, and therefore energy, due
to the advective motion. This radial flow, carrying away thermal
energy from a disc region at a certain $r$ before the expansion
instability has developed at that radius, inhibits the local
thermal energy growing and therefore the disc expansion.

For the case 'b' we show the $r$-$z$ profiles of the disc particles in the two states in figs. 7
and 8. In particular, in fig. 8 we also show the boundaries of the corresponding Shakura-Sunyaev
disc, from which the reader will be able to see the good agreement, in the vertical thickness,
between our simulation and the canonical disc model.

\begin{figure}
\begin{center}
\includegraphics[scale = 0.33,angle = 270.]{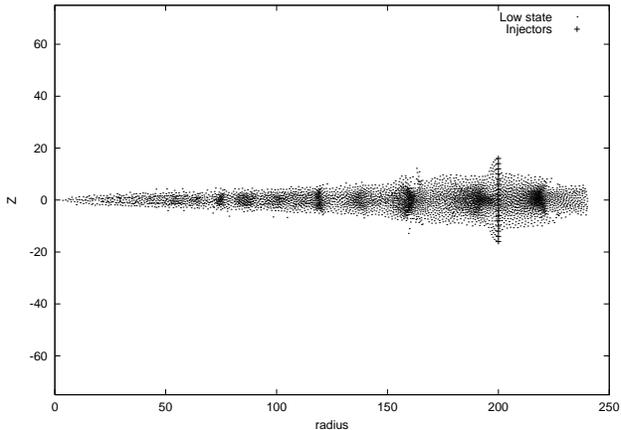}
\caption{The $r$-$z$ profile of the disc of case 'b' in its low
state is shown at the time $t = 0.1016946 \ 10^7 \ R_g/c$ . Every
SPH particle is represented by a small cross. On the $x$ axis the
$r$ values in units of $R_g$ are represented. On the $y$ axis the
$z$ values in units of $R_g$ are represented.}
\end{center}
\end{figure}

\begin{figure}
\begin{center}
\includegraphics[scale = 0.33,angle = 270.]{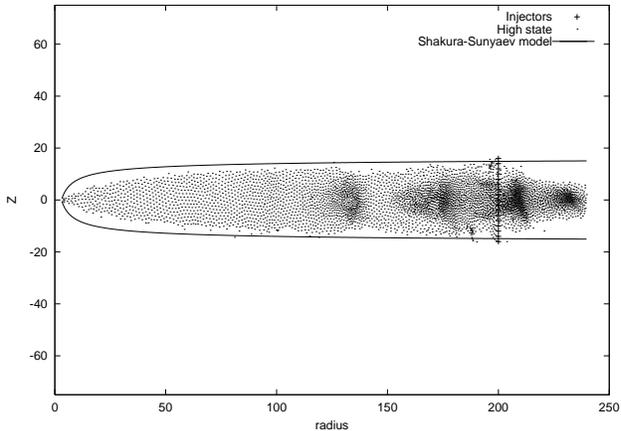}
\caption{The $r$-$z$ profile of the same disc case 'b' in its high
state at the time $t = 0.1167006 \ 10^7 \ R_g/c$ is shown. The
solid line represents the equivalent Shakura-Sunyaev model.}
\end{center}
\end{figure}

The reader will notice that the whole disc is geometrically
thinner in a state than in the other one. The unstable region is
no more a small zone close to the black hole, as in case 'a', but
extends throughout the entire simulation radial range. This is due
to the higher (supercritical) accretion rate, that makes the
radiation pressure dominated zone much wider than in case 'a'. The
extension of the unstable region is also apparent from the
comparison between the radial profiles of the ratio
$P_{rad}/P_{gas}$ in the two states, shown in fig. 9, and from the
similar comparison for the temperature profiles, shown in fig. 10,
where the temperature radial profile (in the disc midplane) of the
corresponding Shakura-Sunyaev model is also included (in order to
show also here the good degree of agreement between simulations
and 1D models).

\begin{figure}
\begin{center}
\includegraphics[scale = 0.33,angle = 270.]{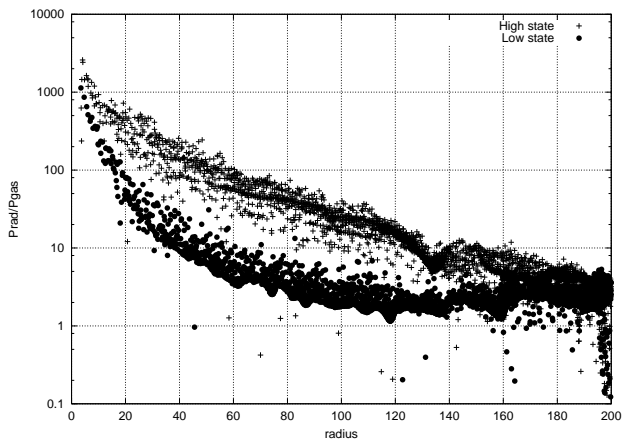}
\caption{The ratio $P_{rad}/P_{gas}$ at the times $t = 0.1016946 \ 10^7$ and $t
= 0.1167006 \ 10^7$ is shown. On the $x$ axis the $r$ values in units of
$R_g$ are represented. The configuration at the earlier time
exhibits a smaller ratio $P_{rad}/P_{gas}$
than the configuration at the later time.}
\end{center}
\end{figure}

\begin{figure}
\begin{center}
\includegraphics[scale = 0.33,angle = 270.]{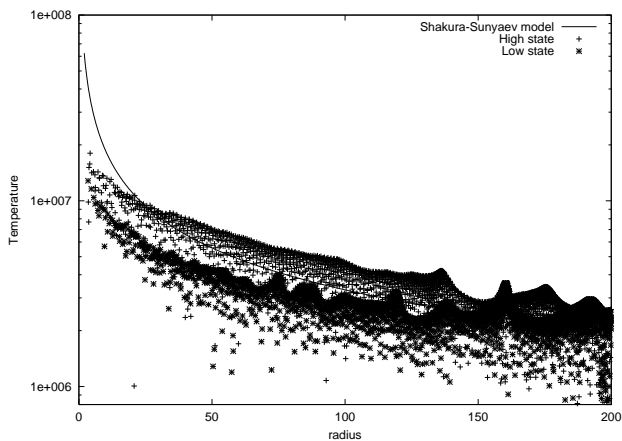}
\caption{The temperature radial profile at the times $t =
0.1016946 \ 10^7$ and $t = 0.1167006 \ 10^7$ is shown. On the $x$
axis the $r$ values in units of $R_g$ are represented. On the $y$
axis the temperature values in kelvin are represented. The
configuration at the earlier time exhibits a smaller temperature
than the configuration at the later time. The solid line
represents the temperature radial profile for the equivalent
Shakura-Sunyaev model.}
\end{center}
\end{figure}

In fact, in both figures the two profiles associated to the two states are
significantly different in the whole disc: approximately up to $200R_g$, where the disc mass
is injected, the geometrically thinner configuration is cooler and less radiation pressure
dominated than the other one (the disc remains however radiation pressure dominated).
So what we said about the 'a' case is also valid in the 'b' one: we can conclude that there is a limit-cycle oscillation between two different disc states, one thinner, cooler and therefore less luminous and the other one thicker, hotter and therefore more luminous. The luminosity
oscillation is shown in fig. 11. In this figure the luminosity time behavior is represented
during the entire formation and evolution of the disc from the time $t=0$, when particles begin
to be injected in a totally empty space.

\begin{figure}
\begin{center}
\includegraphics[scale=0.462,angle=0.]{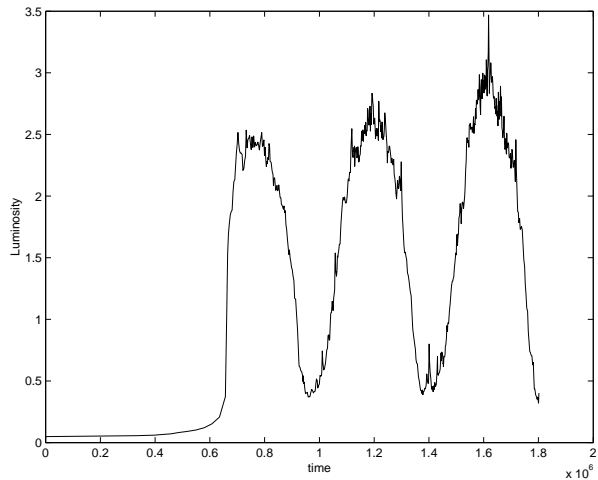}
\caption{The time behavior of the disc luminosity is shown. On the
$x$ axis the time values in units of $R_g/c$ are represented. On
the $y$ axis the luminosity values in units of $L_{theor} = 0.06
\dot{M} c^2 = 2.05 \cdot 10^{39} \hspace{0.20cm} erg
\hspace{0.20cm} sec^{-1}$ are represented.}
\end{center}
\end{figure}

It is evident from this figure the rather different shape of the
time variation curve with respect to the analogous curve for the
'a' case. Here the luminosity $L$ increases approximately as
steeply as it then decreases. So there is no exponential-like
behavior as there is in the 'a' case. We guess that this
difference in the luminosity behavior is due to the much larger
extension of the unstable zone in the 'b' case. In fact, the
cooling of a given disc portion is probably governed by a law
closer to the exponential one if the energy density of the
considered disc region is fundamentally uniform throughout the
region itself. If we indicate with $E(r,t)$ the local energy
density at the radius $r$ and the time $t$ in the disc, the
cooling law at the position $r$ is $\frac{dE(r,t)}{dt} = -k
E(r,t)$, that has for solution the exponential form for $E(r,t)$.\\
It is obvious that this argument holds for the total energy of an entire disc region only
if the variable $E(r,t)$ to consider in the equation above is the same for the whole disc
region, i.e. the energy density $E(r,t)$ is uniform in the considered region.

It is easy to notice that the oscillation shape in the 'b' case is
more similar to the light curves obtained in 1D simulations (see, for
example, \citet{watmin2003}, \citet{Nayakshin} and \citet{SM2001}) than the luminosity
behavior of the 'a' case.
Our hypothesis to explain this fact is that in the 'b' case the
unstable zone is much wider. So the mass to be unloaded by the disc
to pass to the collapsed state is larger. Because of this the refilled
state duration is greater and the oscillation shape shows a luminosity
maximum characterized by a width approximately equal to the minimum
duration. In fact, we have, in the 'b' case, two states with
luminosities different by a factor 7 whose durations are about 25 $s$
each one.\\
In the 'a' case, instead, when the luminosity maximum is reached the
light curve starts immediately its descending (exponential) phase: the
disc holds for a very short time the refilled state, probably because
the disc gets soon rid of the small amount of matter contained in the
small unstable zone.

Finally, we studied the radial behavior of the Mach number $M=\frac{V_r}{v_s}$.

We present the results of this analysis in the figs. 12 and 13,
regarding respectively the collapsed and the refilled state. For
clarity the figures show only the innermost disc region. It is
clear from these figures that the disc has a sonic point,
positioned nearly at $r = 10R_g$ in the collapsed state and (very
approximately) at $r = 15R_g$ in the refilled one. From the
external boundary to the sonic point the radial flow is subsonic,
whereas from the sonic point to the internal boundary we have a
supersonic flow. Though here our data are strongly scattered, we
can say that our 2D simulated models reveal a transonic region
at radii larger than in 1D calculations. For a comparison with
1D models on this aspect of the disc dynamics, one can see, for
example, \citet{SM1998}, where the sonic point is given around
$r = 3R_g$.

\begin{figure}
\begin{center}
\includegraphics[scale = 0.33,angle = 270.]{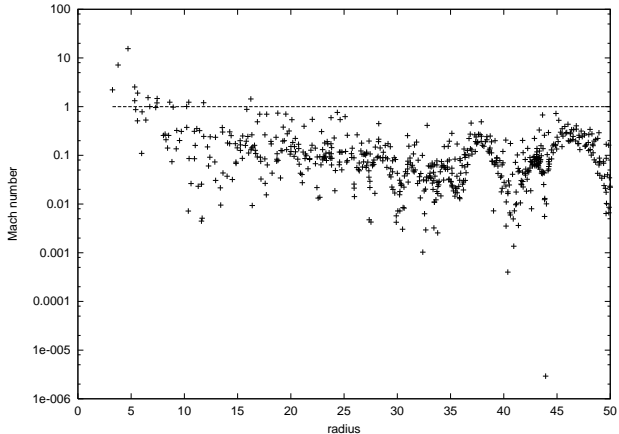}
\caption{The radial profile of the Mach number $M=\frac{V_r}{v_s}$
for the disc of case 'b' in its collapsed state. Every
SPH particle is represented by a small cross. On the $x$ axis the
$r$ values in units of $R_g$ are represented. On the $y$ axis the
Mach number values are represented.}
\end{center}
\end{figure}

\begin{figure}
\begin{center}
\includegraphics[scale = 0.33,angle = 270.]{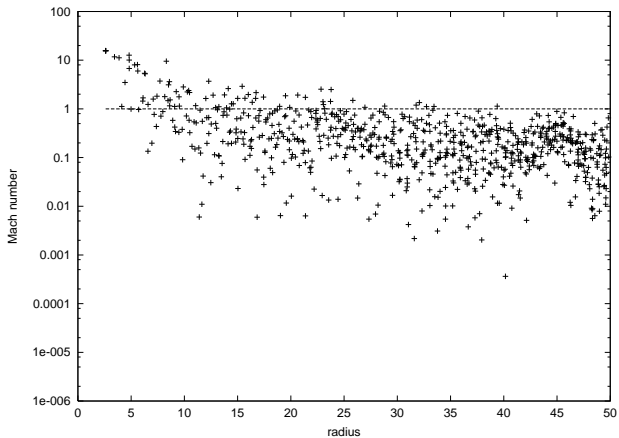}
\caption{The radial profile of the Mach number $M=\frac{V_r}{v_s}$
for the disc of case 'b' in its refilled state. Every
SPH particle is represented by a small cross. On the $x$ axis the
$r$ values in units of $R_g$ are represented. On the $y$ axis the
Mach number values are represented.}
\end{center}
\end{figure}

\section{Discussion}

In this section we discuss three items:

1) confirmation by a true 2D approach of the limit-cycle behavior
produced by the thermal instability;

2) differences between the results obtained by the two approaches
1D and 2D;

3) theoretical considerations about the main features of the
luminosity time behavior: oscillations amplitude, typical times.

1) As we said in the first section, the limit-cycle behavior due
to the Shakura-Sunyaev instability has already been shown as
result of 1D time-dependent simulations
\citep{SM1997,SM1998,SM2001,Janiuk}. In this work we confirm the
existence of the limit-cycle behavior using a 2D approach,
obviously closer to the physical reality of accretion discs.
Moreover, the 2D approach allows to reveal aspects of the
accretion flow that cannot be simulated by the 1D methodology.
Convection and circulation are the main ones. Also, convective
motions are supposed to reduce the thermal instability
\citep{b13}. Just in presence of a significant convective flow,
that our 2D simulations reveal, we confirm the existence of the
thermal instability.

2) In figs. 14 and 15 we show the gas velocity field in a given
disc region for the two disc states. The considered case is the
'b' one.

\begin{figure}
\begin{center}
\includegraphics[scale=0.462,angle=0.]{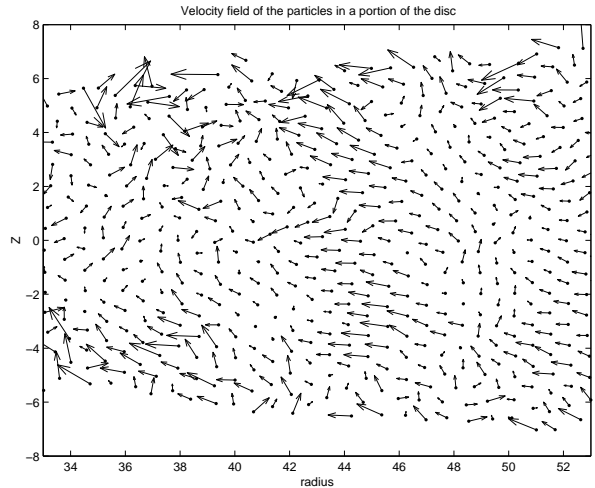}
\caption{The gas velocity field in the radial range between
$33R_g$ and $53R_g$ is shown for case 'b'. The disc state is the
high one. On the $x$ axis the $r$ values in units of $R_g$ are
represented. On the $y$ axis the $z$ values in the same units are
represented. The arrows represent the velocity vectors, with their
lengths proportional to the speed values.}
\end{center}
\end{figure}

\begin{figure}
\begin{center}
\includegraphics[scale=0.462,angle=0.]{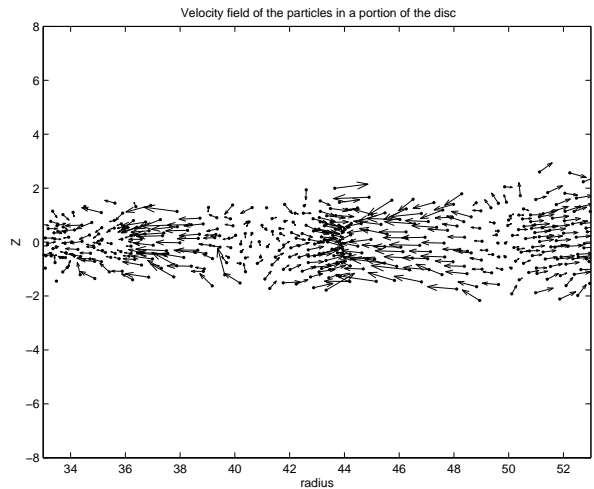}
\caption{The gas velocity field in the radial range between
$33R_g$ and $53R_g$ is shown for case 'b'. The disc state is the
low one. On the $x$ axis the $r$ values in units of $R_g$ are
represented. On the $y$ axis the $z$ values in the same units are
represented. The arrows represent the velocity vectors, with their
lengths proportional to the speed values.}
\end{center}
\end{figure}

These figures make clear the different feature of the two states
(high and low) as regards the convective motions in the disc. From
fig. 14 we can argue the relevant presence of convective and
circulatory motions in the high state, whereas from fig. 15 it is
evident the poor convection in the low state.

These phenomena are important because they affect the time values
that characterize the luminosity behavior. We note that the light
curve we obtain is different from the curves usually obtained by
1D simulations. The luminosity behavior we find is approximately
periodic, as in 1D simulations, but, in case 'a', we obtain
recurrent bursts of time duration not much smaller than the
cycle-time (of about $2$ s), whereas in 1D simulations the bursts
duration is very short in comparison with the cycle-time.

Moreover, convection and circulation also affect the unstable
zone extension: this zone is not the full A zone, extending up
to $30 R_g$, but only a portion of it (up to $13 R_g$),
since the convective motion stabilizes a large fraction of
the radiation pressure dominated region.

In case 'b', the low and high states have about the same duration
(one half of the cycle-time, that is, on average, $50$ s). In 1D
simulations, instead, the high state has an extremely short
duration.

Also, we have compared the disc structures we obtained from our
simulations with the corresponding configurations calculated from
the one-dimensional Shakura-Sunyaev model. The agreement between
the results of the two approaches (the 2D simulated model and the
1D theoretical one) is in general good (see figs. 2 and 4 for case
'a' and 8 and 10 for case 'b'), taken into account that we have a
time varying disc structure. The only significant differences
regard the innermost disc regions: for case 'a', between 10 and 20
$R_g$, the thickness calculated from the canonical model is about
2 times the one of the simulated disc, and also the theoretical 1D
temperature is larger than the simulated one (by 30 $\%$ about);
for case 'b', from 3 to 15 $R_g$, the thickness calculated from
the canonical model is about 2-3 times the one of the simulated
disc, while the theoretical 1D temperature is larger than the
simulated one by 30-35 $\%$ about. So we confirm the result of
Hur\`e and Galliano \citep{hurgall}, who compared vertically
averaged (i.e. 1D) models of accretion discs with the
corresponding vertically explicit (i.e. 2D) configurations and
found that the one-dimensional description of the accretion discs
structure is very close to the two-dimensional one. The generally
good agreement between the disc structures we simulated and the
Shakura-Sunyaev model allows us to conclude that the different
features between the thermal instability we found in 2D and the
limit-cycle behavior obtained by the 1D calculations are not due
to differences in the used disc configurations.

3) There are two time-scales that affect the time features of the
limit-cycle phenomenon: the thermal time-scale, that determines
the development rate of the thermal instability, and the viscous
time, that is connected to the A zone refilling after the collapse
due to the instability. Here we discuss the role of these
time-scales for the case 'a'. In this disc the collapsing zone
extends nearly from $3R_g$ to $13R_g$. We give the values of the
thermal and viscous time-scales at three points in this zone.
These time-scales are quantities calculated from the theoretical
formulae using the values of the disc physical variables resulting
from the simulation. The expression we use for the viscous
time-scale is $t_{visc} =  r^2/\nu$ and, for the thermal time-scale
\citep{b12}:

\begin{equation}
t_{therm} =  \frac{1}{\alpha \Omega}
\frac{A(\beta_r)}{6(5\beta_r-3)}
\end{equation}

where $A(\beta_r) = 8 + 51\beta_r -3\beta_r^2$ and $\beta_r = P_{rad}/
P^{tot}$.\\

Since the viscous time depends on the disc thickness, it assumes
different values in the low (collapsed) state and the high
(refilled) one. So we will distinguish its values in the two
states using the labels 'LS'
(Low State) and 'HS' (High State).\\

\begin{tabular}{|c|c|c|c|}
  \hline
  $r$ & $t_{visc}(LS)$ & $t_{visc}(HS)$& $t_{therm}$ \\
  \hline
  $3R_g$ & $108.9$ s & $7.00$ s & $4.84 \cdot 10^{-3}$ s \\
  $7R_g$ & $2717$ s & $6.79$ s & $2.218 \cdot 10^{-2}$ s \\
  $13R_g$ & $25545$ s & $63.86$ s & $6.046 \cdot 10^{-2}$ s \\

  \hline
\end{tabular}
\\
\\
It is easy to see that the theoretical viscous time-scales are too large compared to
the 'experimental' luminosity cycle-time, whereas the thermal time-scales are too small.

We propose the following explanations. When the disc is in the low
state, the accretion rate in the collapsed zone is small, but, at
the outer boundary of this region, the disc is not collapsed and
has a higher accretion rate. Because of this fact, the accretion
flow at the outer boundary of the collapsed zone 'forces' matter
to enter the region at a rate larger than the rate due to the
viscous time-scale computed inside the region itself. So the
refilling process is accelerated and its time-scale reduced with
respect to the viscous time. A rough estimate of this effect can
be given by the value of the viscous time-scale of the disk just
outside the collapsed zone. Moreover, the calculation of this
viscous time-scale has to take account of the 'real' viscosity
present in the disc: the gas 2D motions (convection and
circulation) give rise to a turbulent flow and consequently to an
additional viscosity, of the order of magnitude $H v_{turb}$
(where $v_{turb}$ is the speed of the turbulent flow), that
has to be added to the Shakura-Sunyaev one.\\
To understand this procedure it must be remembered that, supposing
a viscosity due to a turbulence (the Shakura-Sunyaev $\alpha$-viscosity),
we have not obtained a regular flow with the turbulence 'hidden'
in the $\alpha$-viscosity term. The flow produces another turbulence,
not included in the $\alpha$-viscosity term. Therefore, to express
the total kinematic viscosity, we have to sum the term given by
the speed of this new turbulence to the standard $\alpha$-term.
In formulae, we have a total kinematic viscosity given by:

\begin{equation}
\nu = \alpha v_s H + H v_{turb}
\end{equation}

With this expression for the viscosity, the viscous time-scale
$t_{visc}$ is given by:

\begin{equation}
t_{visc} =  \frac{r^2}{\nu} = \frac{r^2}{\alpha v_s H + H v_{turb}}
\end{equation}

An approximate estimate for $v_{turb}$ can be obtained by adding
the gas radial and vertical speeds. These formulae give for
$t_{visc}$ the value of $12$ s. Finally, it has to be considered
that the transition from the low to the high state is also
accelerated by the process of the radial diffusion of radiation,
whose typical time-scale is $\Delta t = \frac{3 \tau \Delta
l}{c}$, where $\Delta l$ and $\tau$ are the length and the optical
depth of the region through which the radiation diffuses. Taking
account of all these effects, the characteristic time of the
transition LS-HS is lowered from $12$ s to about $5$, value not
far from the luminosity cycle-time we obtain in case 'a' ($2$
s).\\
As regards the inverse process, i.e. the transition from the high
to the low state due to the thermal instability, we form the
hypothesis that the instability development time, that is
essentially the thermal time-scale, is increased by convection
\citep{b13}. Convection is naturally simulated by our 2D code,
whereas the 1D codes, obviously, cannot track the convective
motions of the fluid masses and therefore do not include
convection. This could be the reason for which, in 1D simulations,
the high state duration is very short: the thermal time-scale is
not increased by convection and therefore the instability develops
very rapidly, causing the collapse of the high state in the low
one in a very short time.\\
To support this hypothesis refer to fig. 6 for the radial speed in
the two states low and high. It is clearly shown the extremely low
radial speed of the collapsed zone together with the close
'active' zone exhibiting larger radial speeds. An oscillatory
radial behavior of $V_r$ is also evident.\\
It is interesting that Szuszkiewicz and Miller also found similar
radial speed profiles \citep{SM2001}, with significant
oscillations in the behavior versus $r$. They claim that these
oscillations are a numerical effect and, as a proof of that, show
that, if an artificial diffusion is introduced, the oscillations
disappear. We think, instead, that the radial speed oscillations
are a real physical phenomenon, connected to the gas circulation
in the disc, and that an artificial diffusion can obviously smooth
away the oscillatory behavior, but this is only an artificial
result due to a not physical ingredient.\\
Finally we want to highlight the differences between our results
and Szuszkiewicz and Miller's ones \citep{SM2001} about the hot
gas bulge generated by the thermal instability. In their work this
bulge is just the medium by which the instability propagates
through the disc. When the instability arises, a small region near
the black hole becomes thicker and hotter. Its thickness and
temperature become larger and larger while the radial extension
also increases. The hot gas bulge that is formed through this
process reaches a radial extension of $90R_g$, then it begins to
go down in the region near the black hole. The cooling wave that
accompanies this process propagates gradually towards the internal
boundary, until the whole gas bulge has returned to the disc
equilibrium values of thickness and temperature. In our
simulations the situation is different. The instability starts
with the inner region collapse (therefore with the gas cooling)
and not with the thickness and temperature increasing (the inverse
processes). Both the collapse and the subsequent refilling are
approximately simultaneous over all the unstable region: no
heating and cooling waves propagate through the disc. When the
unstable region is refilled, something similar to the hot bulge of
\citet{SM2001} is formed. The involved region swells, but much
less than in Szuszkiewicz \& Miller's simulation. It reaches a
vertical thickness not much larger than the equilibrium value and,
as said, does not expand radially.

\section{Conclusions}

We put forward evidence with true 2D simulations that the
limit-cycle behavior produced by the Shakura-Sunyaev instability
is present in accretion discs having a radiation pressure
dominated zone (A zone). The time-scale of the instability and the
shape of the light curve depend on the accretion rate. Lower
accretion rates produce shorter time-scales of the oscillations.

The 2D real motions play an important role to calculate the
appropriate values of the oscillation frequencies. We obtain, for
the subcritical regime (case 'a'), a frequency $\nu$ of about 0.5
$Hz$ and, in the supercritical case (the 'b'), $\nu \approx 0.02$
$Hz$. In general the 2D time-scales are shorter (and the
frequencies higher) than the 1D ones. We attribute this result to
the shorter viscous time-scale characteristic of the zone outside
the collapsed region and to the role of the large convection
present in the high state.

These results may be relevant for the explanation of QPO emission
in black hole candidates. More refined models are necessary for a
detailed interpretation of QPO observational data, which, however,
is not the purpose of this paper. Our aim is not to find an
explanation of some sources behavior, but simply to see what
happens to the standard disc structure during its time evolution.

\label{lastpage}


\begin{thebibliography}{99}


\bibitem[\protect\citeauthoryear{Bisnovatyi-Kogan \& Blinnikov}{1977}]{b3}
Bisnovatyi-Kogan G.S., Blinnikov S.I., 1977, A\&A, 59, 111

\bibitem[\protect\citeauthoryear{Brookshaw}{1994}]{brookshaw}
Brookshaw L., 1994, Mem.S.A.It., 65, n. 4, 1033

\bibitem[\protect\citeauthoryear{Chakrabarti \& Molteni}{1993}]{moltchakra}
Chakrabarti S.K., Molteni D., 1993, ApJ, 417, 671

\bibitem[\protect\citeauthoryear{Dilts}{1996}]{Dilts}
Dilts G.A., 1996, Los Alamos National Laboratory Report LA-UR,
96-134

\bibitem[\protect\citeauthoryear{Hubeny}{1990}]{Hubeny}
Hubeny I., 1990, ApJ, 351, 632

\bibitem[\protect\citeauthoryear{Hur\`e \& Galliano}{2001}]{hurgall}
Hur\`e J.-M., Galliano F., 2001, A\&A, 366, 359H

\bibitem[\protect\citeauthoryear{Janiuk et al.}{2002}]{Janiuk}
Janiuk A., Czerny B., Siemiginowska A., 2002, ApJ, 576, 908J

\bibitem[\protect\citeauthoryear{Kippenhahn \& Thomas}{1982}]{Kippen}
Kippenhahn R., Thomas H.-C., 1982, A\&A, 114, 77

\bibitem[\protect\citeauthoryear{Mihalas \& Klein}{1982}]{mihalas}
Mihalas D., Klein R.I., 1982, Jou. Comp. Phys., 46, 97

\bibitem[\protect\citeauthoryear{Molteni et al.}{1998}]{moltbook}
Molteni D., Gerardi G., Valenza M.A., Lanzafame G., Ed. by
Chakrabarti S.K., 1998, Observational Evidence for Black Holes in
the Universe, Kluwer A.P., Dordrecht

\bibitem[\protect\citeauthoryear{Monaghan}{1985}]{b9}
Monaghan J.J., 1985, Comp. Phys. Repts., 3, 71

\bibitem[\protect\citeauthoryear{Monaghan}{1992}]{mon1992}
Monaghan J.J., 1992, Annu. Rev. Astron. Astrophys., 30, 543

\bibitem[\protect\citeauthoryear{Nayakshin et al.}{1985}]{Nayakshin}
Nayakshin S., Rappaport S., Melia F., 2000, ApJ, 535, 798N

\bibitem[\protect\citeauthoryear{Nelson \& Papaloizou}{1994}]{hvar}
Nelson R.P., Papaloizou J.C.B., 1994, MNRAS, 270, 1N

\bibitem[\protect\citeauthoryear{Paczynski \& Wiita}{1980}]{b10}
Paczynski B., Wiita P.J., 1980, A\&A, 88, 23

\bibitem[\protect\citeauthoryear{Shakura \& Sunyaev}{1976}]{b12}
Shakura N.I., Sunyaev R.A., 1976, MNRAS, 175, 613
\bibitem[\protect\citeauthoryear{Shakura et al.}{1978}]{b13}
Shakura N.I., Sunyaev R.A., Zilitinkevich S.S., 1978, A\&A, 62,
179
\bibitem[\protect\citeauthoryear{Shapiro et al.}{1976}]{b14}
Shapiro S.L., Lightman A.L., Eardley D.M., 1976, ApJ, 204, 187

\bibitem[\protect\citeauthoryear{Szuszkiewicz \& Miller}{1997}]{SM1997}
Szuszkiewicz E., Miller J.C., 1997, MNRAS, 287, 165

\bibitem[\protect\citeauthoryear{Szuszkiewicz \& Miller}{1998}]{SM1998}
Szuszkiewicz E., Miller J.C., 1998, MNRAS, 298, 888

\bibitem[\protect\citeauthoryear{Szuszkiewicz \& Miller}{2001}]{SM2001}
Szuszkiewicz E., Miller J.C., 2001, MNRAS, 328, 36

\bibitem[\protect\citeauthoryear{Teresi et al.}{2003}]{ter2003}
Teresi V., Molteni D., Toscano E., 2003, preprint (astro-ph/0307480)

\bibitem[\protect\citeauthoryear{Watarai \& Mineshige}{2003}]{watmin2003}
Watarai K., Mineshige S., 2003, preprint (astro-ph/0306548)


\end{thebibliography}
\end{document}